\documentclass[jcp,twocolumn,superscriptaddress,showpacs]{revtex4}

\usepackage{epsfig}
\usepackage{graphics}
\usepackage{chemfig}
\usepackage{amsmath,bm}
\usepackage{hyperref}
\hypersetup{
 colorlinks = true,
 citecolor= blue,
 linkcolor= blue,
 urlcolor= blue
}
\usepackage{array}

\begin{document}


\title{Quantum simulations of neutral water clusters and singly-charged water cluster anions}

\author{A.~Gij\'{o}n}
\affiliation{Instituto de Ciencia de Materiales de Madrid (ICMM--CSIC), Campus de Cantoblanco,
28049 Madrid, Spain}
\author{E.~R.~Hern\'{a}ndez}
\thanks{email: Eduardo.Hernandez@csic.es} 
\affiliation{Instituto de Ciencia de Materiales de Madrid (ICMM--CSIC), Campus de Cantoblanco,
28049 Madrid, Spain}

\date{\today}

\begin{abstract}
We report a computational study of the structural and energetic properties of water clusters and singly-charged water cluster anions containing from 20 to 573 water molecules. We have used both a classical and a quantum description of the molecular degrees of freedom. Water intra and inter-molecular interactions have been modelled through the SPC/F model, while the water-excess electron interaction has been described via the well-known Turi-Borgis potential. We find that in general the quantum effects of the water degrees of freedom are small, but they do influence the cluster-size at which the excess electron stabilises inside the cluster, which occurs at smaller cluster sizes when quantum effects are taken into consideration.
\end{abstract}


\maketitle

\section{Introduction}
\label{sec:introduction}

Sixty years after its first unequivocal spectroscopic detection~\cite{Hart1962},
the hydrated electron, $e^{-}_{aq}$, continues to fascinate physicists and chemists alike. Since then a great deal of research effort has aimed at gaining a better understanding of its nature and properties, as testified by the large number of review articles devoted to this topic~\cite{Young2012,Turi2012,Herbert2017,Herbert2019}. The reasons for this are several: firstly, $e^{-}_{aq}$ is the simplest reducing agent in solution, and as such it can be expected to play an important role in chemical reactivity in aqueous media. It is a species known to intervene in electron transfer, electrochemistry, radiation effects in water, etc. Its presence has been detected not only in bulk water but also at water-gas interfaces, in thin water layers on metallic substrates, and in water clusters (water cluster anions). One could regard a solvated electron as one of the simplest, if not indeed the simplest, realization of a quantum system interacting with a (quasi)-classical bath, a model system that has been of interest to theoreticians since the early days of quantum mechanics. But perhaps the most obvious reason for the ongoing interest in the hydrated electron is the fact that, in spite of the great efforts, both experimental and theoretical, devoted to understanding this system, there are still several open questions concerning its microscopic nature and properties. At the experimental level, it has proved difficult to correlate spectroscopic signals with a unique and consistent microscopic picture of $e^{-}_{aq}$. At the theoretical level, modelling of $e^{-}_{aq}$ is challenging both because of the need to account for very different degrees of freedom (electron and water molecules), as well as accurately describing their mutual interaction. In spite of these challenges, a widely accepted paradigm of $e^{-}_{aq}$ has emerged over the years, according to which the solvated electron resides in a roughly spherical cavity of excluded volume, being coordinated by approximately four water molecules in the nearest-neighbor shell (see e.g.~\onlinecite{Herbert2019} and references therein). The cavity structural model has been challenged by a non-cavity one, put forward by Larsen {\em et al.\/}~\cite{Larsen2010}. According to this model the solvated electron induces an enhanced water-density region of $\sim$1~nm diameter over which the electron is spread. Both cavity and non-cavity interpretations have been criticised and defended~\cite{TuriComment2011,Herbert2011,Larsen2011,Casey2013}. It should be noted that the one essential difference between the simulations that support either model lies in the way in which the water-excess electron interaction is modelled, a fact that testifies to the difficulty of the problem. 

Experiments and simulations of singly-charged water cluster anions have been frequently employed as a means to gain a better understanding of the properties of $e^{-}_{aq}$, and at the same time have raised interesting questions of their own, such as where does the excess electron reside, inside the cluster or at its surface, how does this depend on cluster size and temperature, or what is the stability range of the cluster. We will be addressing these questions below. 

One of the most common strategies employed to simulate $e^{-}_{aq}$ both in bulk and cluster geometries involves a combination of classical molecular dynamics~(MD) for the water degrees of freedom with a quantum mechanical treatment of the electron, which is generally known as Quantum-Classical~MD. While this may be expected to be a good approximation, it is nevertheless desirable to quantify the effects of treating the water molecules classically, given the well-known importance of nuclear quantum effects in water and aqueous systems \cite{Kuharski1985,Paesani2006,Manolopoulos2009,Vega2010,Vega2011,Richardson2016,Ceriotti2016}, and especially in view of the fact that employing different models for the water-$e^{-}_{aq}$ interaction can lead to structurally different motifs (cavity and non-cavity models). It is also known that the excess electron can occupy either an internal or an external state, depending on such factors as cluster size and temperature; the transition from one type of state to the other may be (and indeed we will show below that it is) affected by the inclusion or otherwise of quantum effects in the molecular dynamics. Fully quantum simulations of water cluster anions have been reported before, such as in the pioneering work of Berne and coworkers~\cite{Thirumalai1986,Wallqvist1987}, and Barnett {\em et al.\/}~\cite{Barnett1987,Barnett1988}, works which treated both water molecules and excess electron within the Path Integral~(PI) formalism. Although feasible, this approach is technically inconvenient, given the disparity of the masses involved: the adequate PI description of the excess electron requires many more beads than the oxygen and hydrogen atoms. An alternative approach is to solve the Schr\"{o}dinger equation for the electron moving in the field resulting from its interaction with the water molecules, and use the PI formalism only for the molecular degrees of freedom. This methodology was, to our knowledge, first used by Takanayagi {\em et al.\/}~\cite{Takayanagi2009} to perform simulations of $(\mbox{H}_2\mbox{O})_5^-$ and $(\mbox{D}_2\mbox{O})_5^-$, and in this sense our work is a continuation of theirs.
Here we report a series of simulations of both neutral and singly-charged water clusters of different sizes and at different temperatures. In order to gauge the relevance of quantum effects in these systems, we perform both classical and quantum simulations of the neutral clusters, and QCMD as well as fully quantum simulations for the water cluster anions. 

The paper is structured as follows: in Sec.~\ref{sec:methods} we provide a succinct description of our computational procedure; full details as well as convergence tests are provided in  Appendix~\ref{sec:appendix}. Our results are presented in Sec.~\ref{sec:results}, focusing first on neutral water clusters, and then on the singly-charged water cluster anions. Finally, we draw our conclusions in Sec.~\ref{sec:conclusions}.

\section{Computational Methodology}
\label{sec:methods}

A detailed description of our computational methodology, including validation and convergence tests is deferred to  Appendix~\ref{sec:appendix}. Here we will confine ourselves to providing a birds-eye view of our simulation procedure.

One of our aims has been to perform both classical and quantum simulations of the equilibrium properties of water clusters and singly-charged water cluster anions, i.e. water clusters incorporating an excess electron. In the latter class of system the excess electron is always treated quantum mechanically by numerical solution of the Schr\"{o}dinger equation for an electron moving in the field generated by its interaction with the water molecules. The molecules were simulated either classically or quantum mechanically within the Feynman Path Integral~(PI) formalism. As is well known, the PI formalism approximately maps the partition function of a quantum system onto that of a classical isomorph, namely a ring polymer with~$P$ monomers, in which each monomer is a classical counterpart of the original system affected by a scaled potential, $V/P$, and coupled to its two nearest neighbors on the ring via harmonic springs. In the limit $P \rightarrow \infty$ one can sample the Boltzmann statistics of the quantum system through its classical isomorph, while in the opposite limit, $P \rightarrow 1$ one falls back onto the classical description. Convergence with respect to the value of $P$ used in the simulations of the quantum systems has been monitored and is reported in Appendix~\ref{sec:appendix}.

The water intra- and inter-molecular interactions are described by means of the SPC/F model~\cite{Toukan1985,Lobaugh1997}. Harmonic springs account for the O-H bond and H-O-H bond angle dynamics; the intermolecular interactions consist of a Coulomb contribution between partial charges located at the atoms ($q_{\text{O}} = - 0.82 e = -2 q_{\text{H}}$, where $e$ is the quantum of charge), and a Lennard-Jones potential acting between oxygen atoms. 

In cluster anions the excess electron interaction with the oxygen and hydrogen atoms of water molecules is described via the Turi-Borgis~\cite{Turi2001,Turi2002} local pseudopotential, which is parametrized to reproduce the ground-state energy and the optical absorption spectrum of the hydrated electron with classical MD simulations of bulk water at 298\,K. In the limit of long distances this model reproduces the Coulomb interaction between electron and atomic partial charges, while at short distances it mimics the interaction of the electron with corresponding neutral atoms, thus avoiding the Coulomb divergence. The model also includes a term accounting for the induced polarization of oxygen atoms brought about by the presence of the excess electron. In total, the potential energy accounting for the electron-water interaction has the following expression:
\begin{equation}
  \begin{split}
    V_{e-\text{W}}({\bf r}) {} =  \sum_{n=1}^{N_\text{W}}\Big[ V_{e-\text{O}}\big( r^{(n)}_{\text{O}} \big) +
    V_{\text{pol}}\big( r^{(n)}_{\text{O}} \big) + \\
    {} V_{e-\text{H}} \big(r^{(n)}_{\text{H}_1} \big) + 
    V_{e-\text{H}} \big(r^{(n)}_{\text{H}_2} \big) \Big],
  \end{split}
\label{eq:w-e-potential}
\end{equation}
where $r^{(n)}_{i} = |{\bf r} - {\bf r}^{(n)}_{i}|$, and ${\bf r}^{(n)}_{i}$ is the position of atom $i$ in molecule {\em n\/}; the sum extends over all molecules in the cluster. The explicit form of the terms appearing in Eq.~(\ref{eq:w-e-potential}) can be found in refs.~\onlinecite{Turi2001,Turi2002}.

We used standard Molecular Dynamics~(MD) to sample the thermal properties of either the classical systems or their quantum counterparts, using a time step of 1~fs and coupling a Bussi-Parrinello~\cite{Bussi2007} thermostat to each degree of freedom in order to ensure appropriate sampling of the canonical ensemble. 

\section{Results and discussion}
\label{sec:results}

In this section we report the results of the classical and quantum molecular dynamics simulations of neutral and negatively charged water clusters with $n$ from 20 to 573 and temperatures from 50 to 400\,K. First, in~\ref{sub:neutral}, we consider the case of neutral clusters, focusing in particular on the observed differences in energetic and structural equilibrium properties between the classical and quantum treatment. Then, in~\ref{sub:charged} we report the results obtained for clusters charged with an excess electron. 

\subsection{Neutral water clusters}
\label{sub:neutral}

We have computed the energetic and structural properties of neutral water clusters
$(\text{H}_2\text{O})_{n}$ with $n$ in the range 20 to 237, both by means of classical and PI MD simulations at temperatures between 50 and 400~K. 

\begin{figure}[t]
\centering 
\includegraphics[scale=0.5]{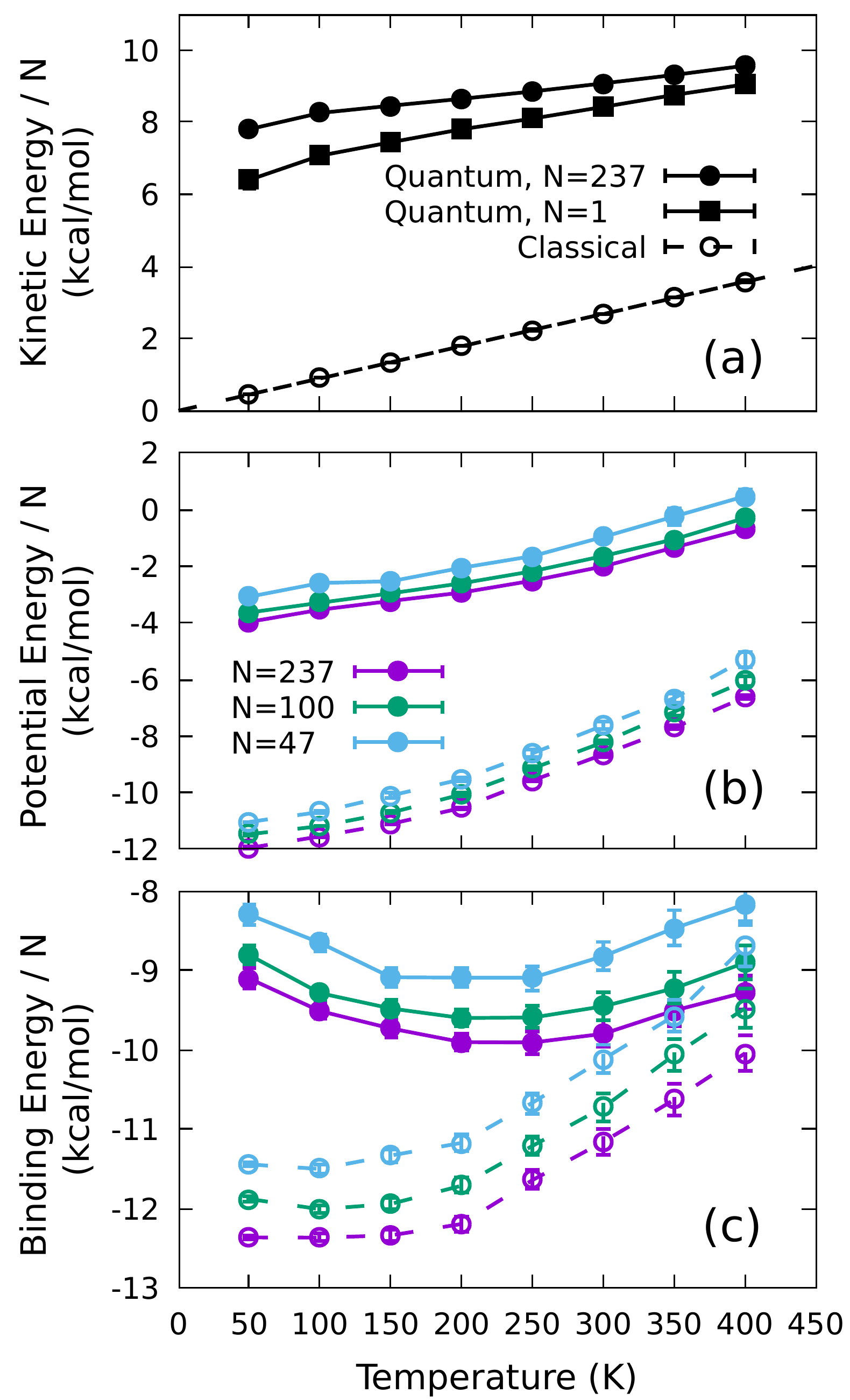}
\caption{Averages of the kinetic energy (a), potential energy (b) and binding energy (c) per water molecule as a function of temperature, for quantum (solid line and filled symbols) and classical (dashed line and empty symbols) simulations of neutral $(\text{H}_2\text{O})_{N}$ clusters.}
\label{fig:neutral-energy}
\end{figure}

Let us first consider the dependence of the energetic properties on temperature for some representative cluster sizes, as shown in Fig.~(\ref{fig:neutral-energy}). Panel (a) displays the kinetic energy per molecule for $n=1$ and $n=237$. It can be seen that in the classical treatment the kinetic energy is independent of cluster size, displaying a linear temperature behavior, as expected. In contrast, the quantum results do show a small but noticeable size dependence, with the results for intermediate cluster sizes (not shown) falling between the curves for $n=1$ and 237. It can also be seen that, while the classical results extrapolate to zero at $T = 0K$, this is not so for the quantum results, which tend to a size-dependent constant value, namely the zero-point energy~(ZPE). In panel (b) of Fig.~(\ref{fig:neutral-energy}) we show the potential energy, again as obtained from both classical and quantum simulations, for sizes $n=47,100,237$. The difference in average potential energy between the quantum and classical results for a given size and temperature corresponds closely to the ZPE, implying that the quantum clusters are predicted to be somewhat less energetically stable than their classical counterparts, at least at low temperatures.  

\begin{figure}[t]
\centering 
\includegraphics[scale=0.55]{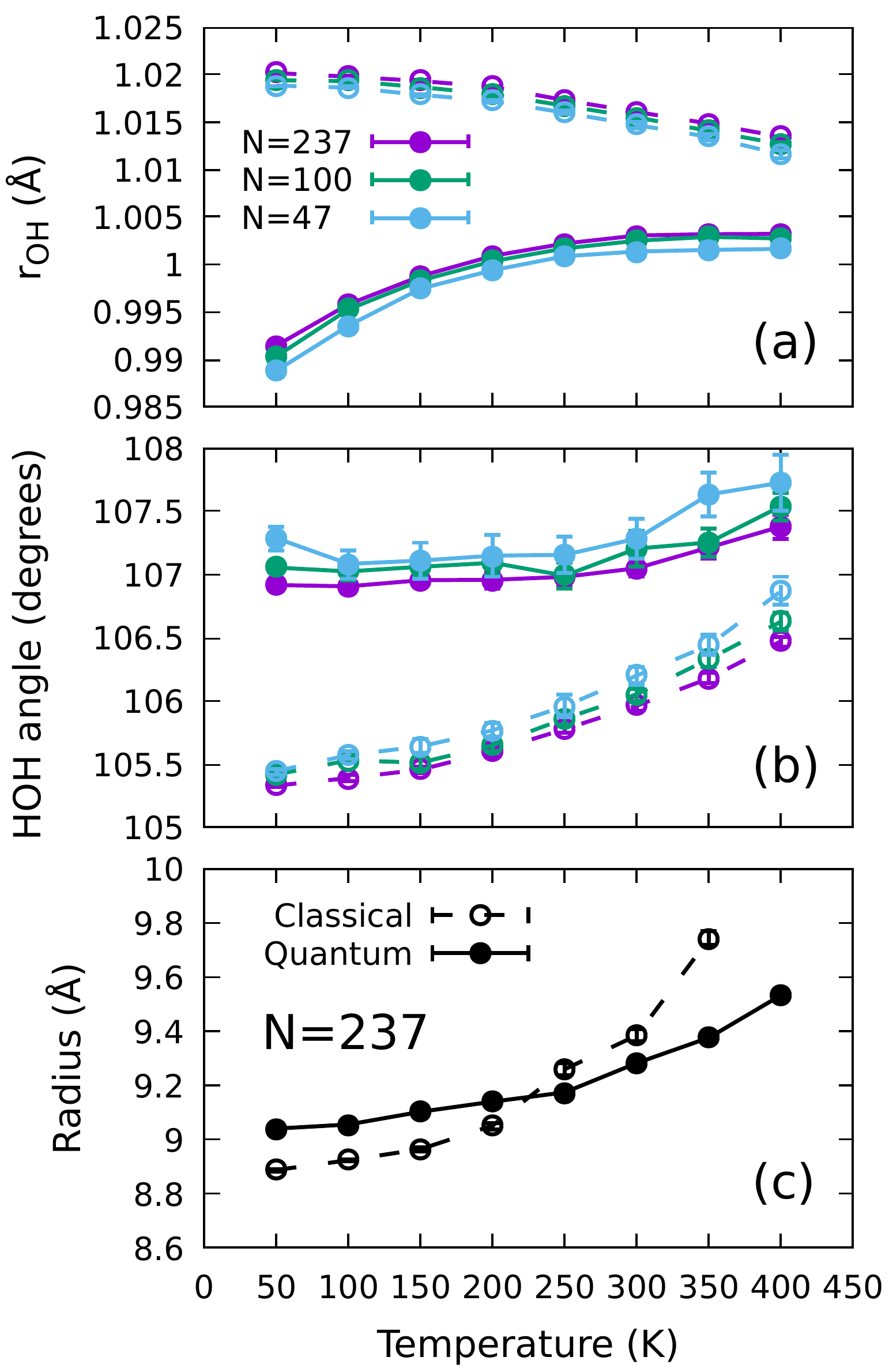}
\caption{Averages of the intra-molecular O-H distance (a) and H-O-H angle (b), and cluster radius (c) of neutral $(\text{H}_2\text{O})_{N}$ clusters, as a function of the temperature, for quantum (solid line and filled symbols) and classical (dashed line and empty symbols) cases.}
\label{fig:neutral-structure}
\end{figure}

The formation or binding energy of a neutral cluster is defined as
\begin{equation}
\Delta E_n = E[(\text{H}_2\text{O})_{n} ] - n E(\text{H}_2\text{O}) \,,
\label{eq:binding-energy}
\end{equation}
where $E(\text{H}_2\text{O})$ is the average total energy of a single water molecule. Binding energies per molecule are shown in Fig.~(\ref{fig:neutral-energy}c). As can be noted, the classical binding energies are always lower (higher in absolute value) than the quantum ones for the same temperature and cluster size, testifying to the lower stability of the latter. Although the classical and quantum curves approach one another at higher temperatures, the differences are appreciable in all the temperature range considered here. We note that the quantum binding energy decreases (increases in absolute value) at temperatures between 50 and 250\,K, even if the quantum potential energy  [see Fig.~(\ref{fig:neutral-energy}b)] is monotonically increasing at all temperatures. In contrast, the classical binding energy remains essentially constant up to 150\,K, increasing linearly above 200\,K.

Next we consider some structural properties of $(\text{H}_2\text{O})_n$ clusters. Fig.~(\ref{fig:neutral-structure}-a) and (b) display the calculated temperature dependence of intra-molecular O-H distances and H-O-H bond angles. As can be seen, both the values as well as their temperature behavior are different in the classical and quantum results. Although the differences are not large, they are nevertheless appreciable; it is particularly noticeable that the temperature dependence of the O-H distance displays opposite trends, decreasing for the classical results, while increasing in the quantum simulations. The bond-angle temperature behavior is increasing in both cases, although the average values are almost two degrees larger in the quantum case at low temperatures. The results of both types of simulations can be seen to converge to the same values in the limit of high temperatures, as expected, but differences are still appreciable at the highest temperature considered here, namely 400~K. 

In order to quantify the size of clusters at different temperatures we calculate their gyration radius, shown in Fig.~(\ref{fig:neutral-structure}-c) for the particular case $n = 237$. At low temperatures the effective gyration radius of the quantum cluster is larger than the classical one, which is consistent with the higher degree of delocalization and lower stability in the quantum counterpart. Interestingly, however this trend is reversed at temperatures near and above 250~K. 
We attribute this change in the classical results to a solid-liquid phase transition. Indeed, this seems to be corroborated by the fact that the slope of the mean squared displacements of the water molecules in the classical system increase by nearly an order of magnitude upon raising the temperature from 200 to 250~K, giving a diffusion coefficient of $\sim 4.3 \times 10^{-9} \mbox{m}^2/\mbox{s}$, a value comparable to that obtained from simulations of super-cooled water at the same temperature~\cite{Park2015}. 
\begin{figure}[t]
\centering 
\includegraphics[scale=0.36]{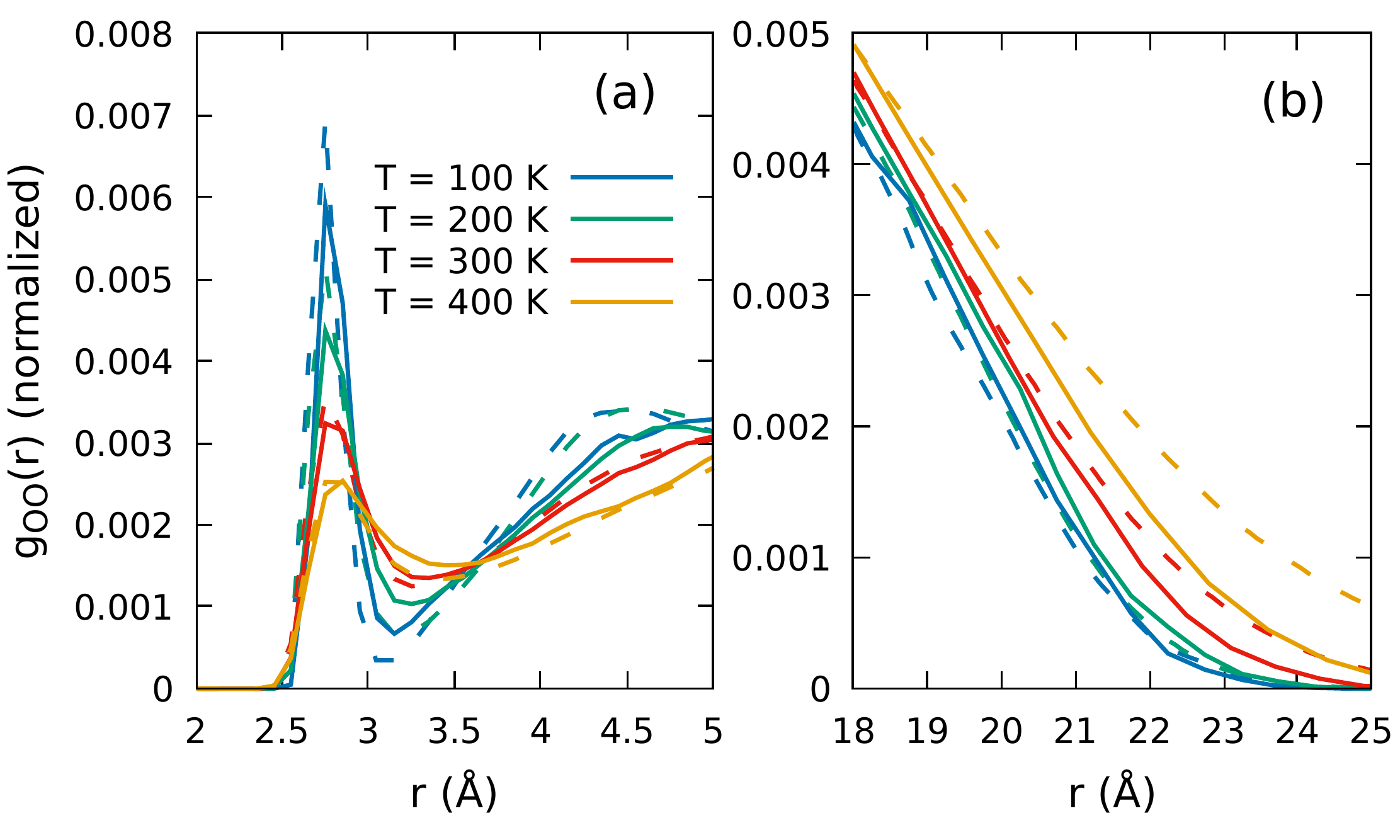}
\caption{Radial pair distribution function of the oxygen-oxygen ($g_{\text{OO}}$) distance for quantum (solid line) and classical (dashed line) simulations of the neutral $(\text{H}_2\text{O})_{237}$ cluster.}
\label{fig:neutral-gOO}
\end{figure}

Quantum and thermal effects are also observable in the pair radial distribution functions (RDFs). In Fig.~(\ref{fig:neutral-gOO}) we report the short- and long-range regions of the oxygen-oxygen RDF, $g_\text{OO}$(r), of the $n=237$ cluster for the classical and quantum cases at temperatures ranging from 100 to 400\,K. The first peak of the $g_\text{OO}$ function at 100 and 200\,K [see panel (a)] at  $r \approx 2.75$\,\AA\ and the second at $r \approx 4.5$\,\AA\ correspond to the first and second neighbor oxygen-oxygen distances in ice, respectively. As temperature is increased the intensity of the peaks reduces and they become broader; also the minima located between the first- and second-nearest neighbour peaks become gradually filled. Likewise, higher temperatures result in more extended tails at long distances, as seen in panel (b). These effects are present in both the quantum and classical simulations; however, if we compare the quantum and classical results in more detail, we can see that at short distances the quantum analysis results in broader peaks of lower intensity, an effect that is particularly noticeable at lower temperatures. This broadening is to some extent equivalent to an increased temperature effect. At higher temperatures, the distinction is more apparent in the long-distance tails of the RDF distributions (panel b), where the classical clusters are seen to be slightly more extended, which is consistent with a greater effective radius. 

Thus, taking into account the previous analysis of the RDFs and that of the cluster effective radius seen in Fig.~(\ref{fig:neutral-structure}-c), we can conclude that the classical description leads to more highly structured and compact clusters at temperatures below 200\,K, a general conclusion that is in agreement with previous work~\cite{Kuharski1985,Wallqvist1985}, whereas the reverse is true for temperatures above 200\,K. 

\subsection{Water cluster anions}
\label{sub:charged}

Now we turn to consider the properties of singly-charged water cluster anions, $(\text{H}_2\text{O})_n^{-}$; specifically, we report a series of classical and quantum simulations with $n=20,32,47,76,100,139,237,573$ at temperatures of 100, 200 and 300\,K.

\begin{figure}[t]
\centering
\includegraphics[scale=0.32]{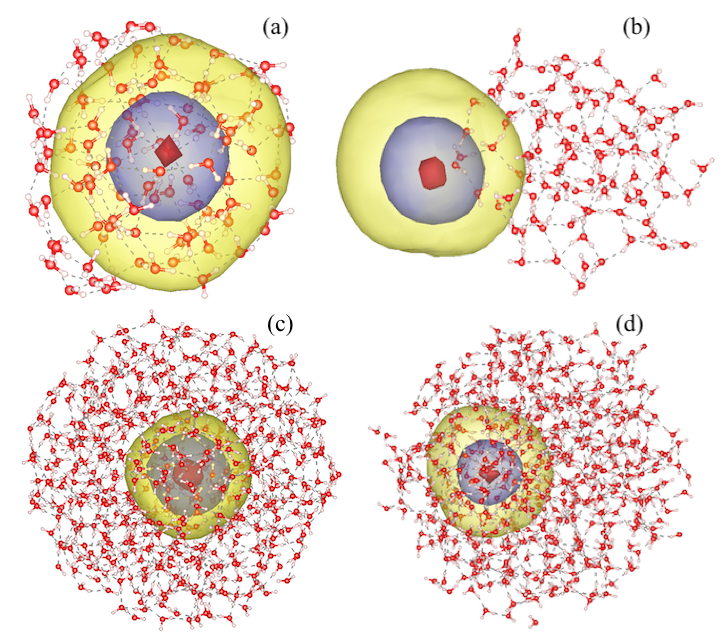}
\caption{Representative configurations of water cluster ions obtained from quantum simulations. Water molecules are shown at their centroid positions; the electron charge density is represented as concentric spheres, the outer one containing $\sim$\,90\,\%\ of the charge density, the intermediate one $\sim$\,50\,\%, and the inner one $\sim$\,2\,\%. (a) 100 molecules, T = 100~K; (b) 100 molecules, T = 300~K; (c) 573 molecules, T = 100~K; (d) 573 molecules, T = 300~K.}
\label{fig:sample-configurations}
\end{figure}

In Fig.~(\ref{fig:sample-configurations}) we show some representative configurations obtained from the quantum simulations of cluster anions. Panels (a) and (b) show two configurations of the $(\mbox{H}_2\mbox{O})^-_{100}$ anion at 100 and 300~K, respectively. As can be seen, the electron delocalises over a volume that is comparable to that of the entire cluster. At low temperature the excess electron is accommodated inside the cluster, but at 300~K it is attached to its surface. At 200~K (not shown) the electron remains inside the cluster, but locates itself closer to the surface. Larger cluster sizes seem to stabilize the electron inside, as shown in panels (c) and (d), where we display instantaneous configurations of $(\mbox{H}_2\mbox{O})^-_{573}$ at 100 and 300~K. It should be noted that at the higher temperature these structures are probably unstable; during the course of our simulations we see that clusters of size $n \leq 100$ occasionally evaporate molecules form the surface at 300~K; the same probably happens to larger clusters over longer periods of time, but all cluster sizes remain intact over the course of our simulations for temperatures of 200~K and lower. For cluster anions larger than $(\mbox{H}_2\mbox{O})^-_{100}$ the electron is always observed to remain inside the cluster, but moving off-center towards the surface, as can be appreciated in panel (d). 

Throughout each simulation, we compute the average distance between the electron mean position and the cluster's center of mass, $d_e=\sqrt{\langle|\bar{\bm{r}}|^2\rangle}$, with $\bar{\bm{r}}$ being the mean position of the electron density. As mentioned in the introduction section, and seen in Fig.~(\ref{fig:sample-configurations}), the excess electron 
can be found either in an interior state (within the cluster) or attached to the surface, depending on factors such as cluster size and temperature. In order to classify the electron as being in one or the other state, it is helpful to adopt the following criteria:
\begin{align}\label{eq:criteria}
\begin{split}
 d_e+R_e<R_c & \implies \text{Interior state} \,, \\ d_e+R_e \gtrsim R_c
 & \implies \text{Surface state} \,,
\end{split}
\end{align}
where $R_\text{c}$ is the gyration radius of the cluster and $R_e$ is that for the electron [see Eq.~(\ref{eq:egyration})]. The quantity $d_e + R_e$ can thus be interpreted as the outer reach of the electron, that is, the maximum distance from the cluster's center of mass where the electron density has an appreciable value.

%
\onecolumngrid
\begin{center}
\begin{figure}[t]
\includegraphics[scale=1.1]{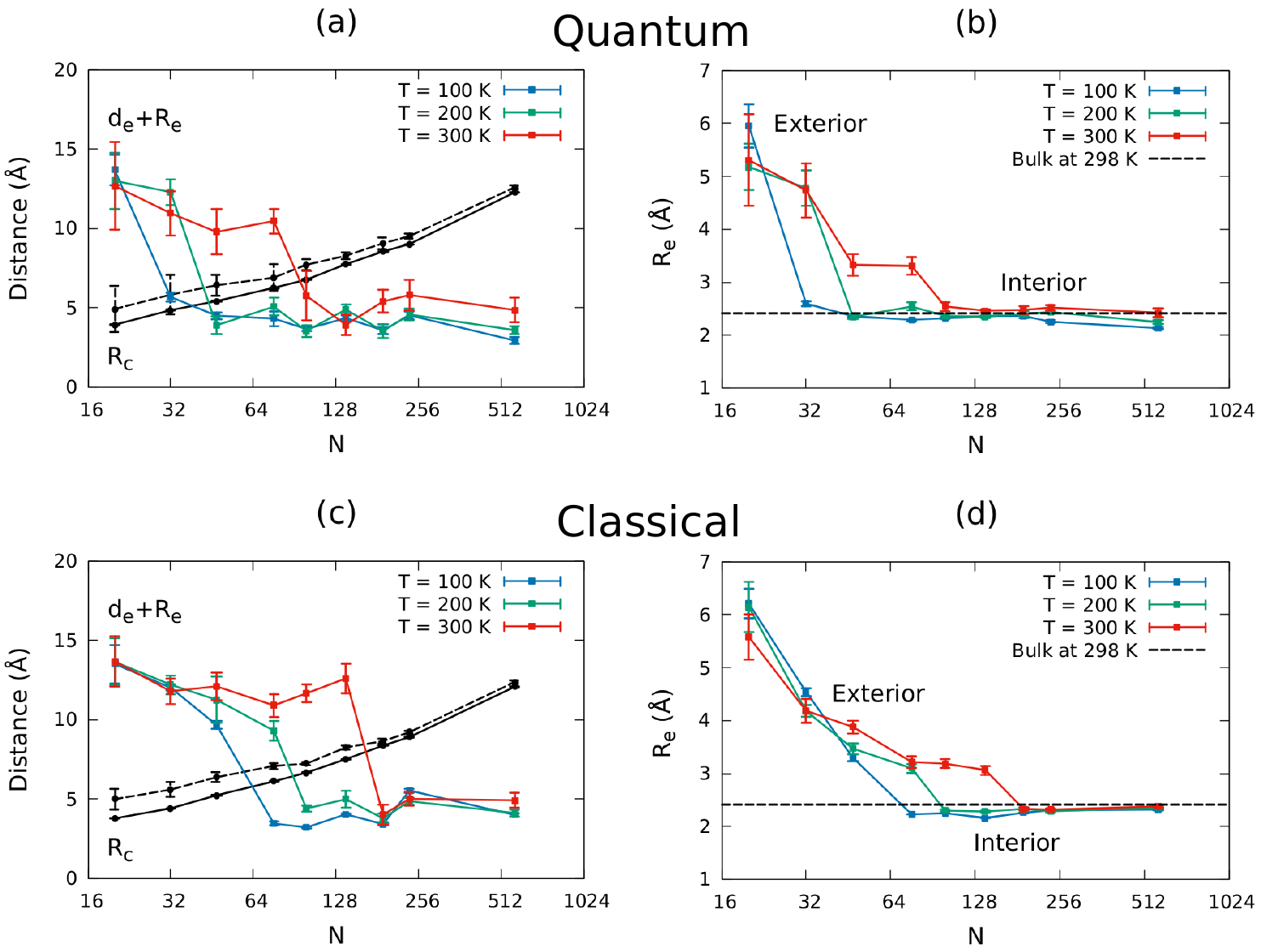}
\caption{(a) Comparison of the excess electron reach, $d_e+R_e$ and cluster radius, $R_c$, as a function of the cluster size for different temperatures. The solid and dashed lines represent $R_c$ at 100 and 300\,K of temperature, respectively. (b) Gyration radius of the excess electron (in color) along with the simulated value for bulk water at 298\,K (dashed black line) of Ref.~\onlinecite{Turi2002}. Panels (c) and (d) show the same quantities as (a) and (b) for the classical case.}
\label{fig:charged-ereach}
\end{figure}
\end{center}
\twocolumngrid
%

Concerning differences between the fully quantum and quantum-classical treatment of the cluster anions,
we can say that in general we obtain qualitatively similar trends for the excess electron properties, but some quantitative differences can be identified. First, we classify the electronic states as interior or surface states, following the criteria of Eq.~(\ref{eq:criteria}). In panels (a) and (c) of Fig.~(\ref{fig:charged-ereach}), we display the sum $d_e+R_e$, which, as argued above gives an indication of the outer reach of the excess electron along the cluster radius, as a function of the cluster size. The crossing of  $d_e+R_e$ and the cluster radius indicates a transition from an exterior to an interior state. For classical simulations of Fig.~(\ref{fig:charged-ereach}-c), small clusters ($n\leq47$) can only accommodate an exterior state at all temperatures, while bigger clusters ($n\geq190$) can host the excess electron in an interior configuration. The transition from surface to interior states occurs at different sizes depending on the temperature. In general, when increasing the temperature, the transition takes place at larger sizes. At 100, 200 and 300\,K, the transitions appear for $n$ in the ranges 47-76, 76-100 and 139-190, respectively. Outside the transition region of intermediate sizes of $n=47-190$, the differences between temperatures are very small. For the quantum simulations, for which results are plotted in Fig.~(\ref{fig:charged-ereach}-a), the most remarkable difference with the classical results is that the transition from surface to interior states occurs at smaller cluster sizes. At 100, 200 and 300\,K, the discontinuity appears for $n$ in the ranges 20-32, 32-47 and 100-139. We attribute this to the fact that the quantum simulations allow atoms to explore classically forbidden regions, thus increasing the possibility of stabilization of the excess electron into the more strongly bound interior state. We should point out that the results plotted in Fig.~(\ref{fig:charged-ereach}) should be taken as qualitative trends, because the criteria of Eq.~(\ref{eq:criteria}) rely on the assumption that the cluster geometry is spherical, which may not be the case for small clusters and/or high temperatures. Nevertheless the observed trends appear to be robust.

The excess electron gyration radius, which measures its dispersion around its mean position, is also characteristic of each type of configuration. In panels (b) and (d) of Fig.~(\ref{fig:charged-ereach}) we see how the gyration radius evolves towards the bulk value of 2.42\,\AA\, obtained by Turi and Borgis for water at 298\,K in Ref~\onlinecite{Turi2002}, with the same model used here. This value agrees well with the experimental moment analysis of the absorption spectrum~\cite{Lavrich1994,Bartels2001}, which results in a radius of 2.5\,\AA\ in bulk water. As can be observed, the gyration radii for exterior configurations approach the bulk value from above as the cluster grows, whereas interior states have a nearly constant value, regardless the temperature or the cluster size.  
\begin{figure}[t]
\centering 
\includegraphics[scale=2.0]{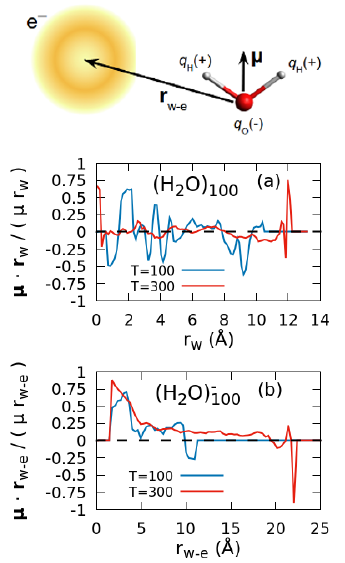}
\caption{Projection of the dipolar moment of water molecules for the $n=100$ neutral cluster (a) and in the presence of the excess electron (b). For the neutral cluster, the projection is on the radial direction, that is, the line between the center of mass of the cluster and each specific water molecule. For the negatively charged  cluster, the dipolar moment is projected on the direction linking the mean position of the excess electron with each water molecule.}
\label{fig:charged-dipoles}
\end{figure}

In order to gain insight into the mechanisms resulting in the different configurations of the excess electron, in Fig.~(\ref{fig:charged-dipoles}) we plot the projection of water molecule dipoles onto the cluster radial direction vs distance from the cluster center of mass for a neutral cluster (panel a),  and onto the axis separating water molecule and electron center of mass for the cluster anion of the same size (panel b), at two different temperatures. In the SPC/F water model the molecular dipole lies close to the H-O-H angle bisector, pointing in the direction of the hydrogen atoms. Firstly, let us consider the orientation of water molecule dipoles in the neutral $n=100$ cluster, in Fig.~(\ref{fig:charged-dipoles}-a). At $T=100$\,K the dipole projection onto the cluster radial direction displays an oscillating character consistent with the interpretation of the cluster being solid. Indeed, at 300~K the oscillating character has been considerably washed out, although still appreciable to some extent, indicating that the cluster is liquid. It can be seen that in both cases the outer layer of the cluster has molecules preferentially oriented such that the dipole, i.e. the hydrogen atoms, point away from the cluster.
In the case of the cluster anions, Fig.~(\ref{fig:charged-dipoles}b), the nearest water molecules to the electron are oriented such that their dipoles point towards the electron center of mass, as indicated by the positive values of the projection in the direction $\bm{r}_{\text{w-}e} =  \bar{\bm{r}}_e - \bm{r}_{\text{w}}$ at short distances. This is the case for both internal and external electronic states. 

\begin{figure}[tb]
\centering 
\includegraphics[scale=1.5]{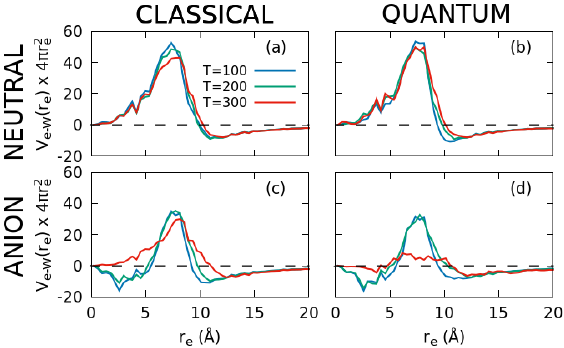}
\caption{Radial electron potential from the center of the water cluster for the classical neutral (a), quantum neutral (b), classical charged (c) and quantum charged (d) clusters with $n=100$ water molecules.} 
\label{fig:charged-rpot}
\end{figure}

The orientation of the molecular dipoles in the outer layer of the neutral clusters, see Fig.~(\ref{fig:charged-dipoles}-a), generates a surface dipole that can serve as a trap for an excess electron. This is shown by the radial electronic potential profiles of panel (a) and (b) of Fig.~(\ref{fig:charged-rpot}), obtained from classical and quantum simulations, respectively. Both exhibit a potential energy well close to the cluster surface. In the presence of the excess electron (panels c and d), there is a distortion of the molecular network that can generate a stable region inside the cluster, at shorter distances from the cluster center; this is akin to the well-known {\em polaron effect\/} in extended systems. For the classical simulation of $(\text{H}_2\text{O})_{100}^{-}$, see Fig.~(\ref{fig:charged-rpot}c), an interior state is only stable at temperatures of 100 and 200\,K, but becomes metastable at 300\,K. In the quantum case (panel d) the situation is similar, though the internal state remains slightly more stable as the temperature is increased than in the classical case. In contrast, the external potential trap seems to be more robust against temperature increase: the surface potential minimum becomes shallower as the temperature is increased, but is less affected and remains attractive at all temperatures considered here. Comparison between the results obtained from classical and quantum simulations [Fig.~(\ref{fig:charged-rpot}c) and (d)] indicate that in the latter case the interior state is somewhat more robust against temperature increase, which is consistent with our earlier observations [see Fig.~(\ref{fig:charged-ereach})].

\begin{figure}[t]
\centering 
\includegraphics[scale=1.0]{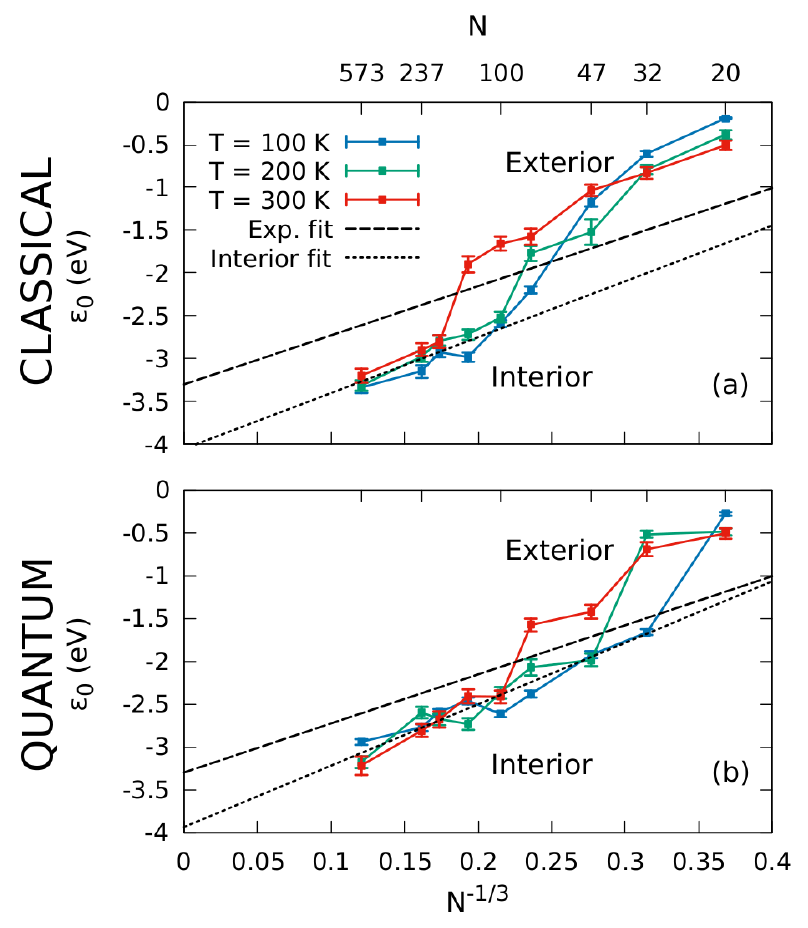}
\caption{Classical (a) and quantum (b) average of the simulated ground-state energy of the excess electron for $(\text{H}_2\text{O})_n^{-}$ clusters at different temperatures. The experimental fit of Coe from Ref.~\onlinecite{Coe1990} is shown with dashed black lines, while the linear fits from our interior-state data are represented with dotted black lines.}
\label{fig:charged-E0}
\end{figure}

Let us now focus on the energetic properties of the excess electron. Its average ground state energy, $\varepsilon_0$, can be compared with the negative of the experimental vertical detachment energy~(VDE) as obtained via photoelectron spectroscopy~\cite{Coe1990,Verlet2005,Issendorff2009}. The VDE is expected to have a linear dependence with the inverse cluster radius (or equivalently, $n^{-1/3}$) for both surface or interior states~\cite{Makov1994}. This was indeed observed in the experimental results of Coe {\em et al.\/}~\cite{Coe1990}, whose linear fit is shown in Fig.~(\ref{fig:charged-E0}) as a dashed black line. Our average values for $\varepsilon_0$ for surface states appear above the experimental line, while those of internal states are found below it, in agreement with the simulations of Turi {\em et al.\/}~\cite{Turi2005}. This is the case for both the classical (a) and quantum (b) simulations, the only difference being that the latter evidence a transition to the internal state at smaller cluster sizes, consistent with our earlier discussion [see Fig.~(\ref{fig:charged-ereach})]. Extrapolation of our calculated average ground state energies for the interior states to the bulk limit results in an energy close to $-4$\,eV, somewhat lower than the value of -3.1\,eV found by Turi  {\em et al.\/}~\cite{Turi2005}, which is closer to the experimental value of -3.3\,eV~\cite{Siefermann2010,Tang2010}. This discrepancy could at least in part arise from the fact that the model we are using was originally designed to describe the excess electron in bulk water, and has not been readjusted for the case of cluster geometries. Out of the classical and quantum results, the latter are somewhat closer to the experimental value, but still below it. 

\begin{figure}[t]
\centering 
\includegraphics[scale=1.0]{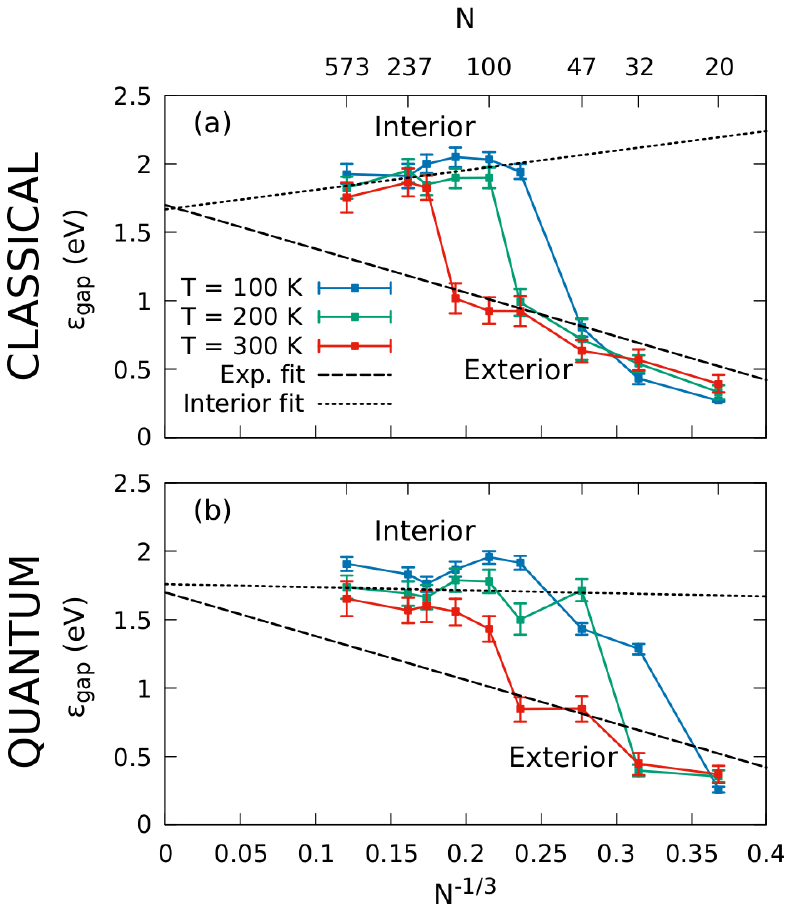}
\caption{Classical (a) and quantum (b) average of the simulated energy gap of the excess electron for $(\text{H}_2\text{O})_n^{-}$ clusters at different temperatures. The experimental fit of Ayotte from Ref.~\onlinecite{Ayotte1997} is shown with dashed black lines, while the linear fits from our interior-state data are represented with dotted black lines.}
\label{fig:charged-Egap}
\end{figure}

In Fig.~(\ref{fig:charged-Egap}) the calculated energy gaps between first excited and  ground state, $\varepsilon_{\text{gap}} = \varepsilon_1 - \varepsilon_0$, are shown together with the experimental fit obtained from the maxima of the optical adsorption spectra by Ayotte {\em et al.\/}~\cite{Ayotte1997}. It is evident from the figure that only the external-configuration states agree with the experimental fit, which indicates that in the experiments of Ayotte {\em et al.\/}, conducted in cluster sizes in the range 6 to 50, only surface states were observed. It is noteworthy that the extrapolations of the linear fit to the calculated band-gap for both interior and  exterior states to infinite size fall very close to the experimentally observed value of 1.72\,eV at 298\,K~\cite{Jou1977,Jou1979}.

The most remarkable difference between the classical and quantum simulations is the appearance of a transition from surface to internal states at smaller sizes in the latter case. This trend is observed at all temperatures considered here. Specifically, at 100\,K, no interior configurations of the excess electron are observed for clusters smaller than $n=76$, with the transition occurring somewhere between 47 and 76. In the quantum simulations, however, the transition appears to occur in the range $n=20-32$. This observation is in good agreement with the conclusions of Neumark~\cite{Neumark2008}, who carefully analyzed the experimental VDE data  previously assigned to internal states~\cite{Verlet2005}, concluding that the transition occurs in the size range $n=25-35$. Furthermore, this assessment is in full agreement with other low temperature experiments \cite{Issendorff2009,Herburger2019} and consistent with theoretical works \cite{Sommerfeld2006,Frigato2008}, which observe internal states for clusters with $n<50$.

\section{Conclusions}
\label{sec:conclusions}

To summarize, we have reported results of simulations of neutral water clusters and cluster anions containing an excess electron, with cluster sizes in the range 20 to 573 molecules. We have treated the water molecules both classically and quantum mechanically via the Path Integral formalism. Our results indicate that quantum effects in the molecular degrees of freedom generally have a small but noticeable influence on the cluster properties, and in particular they affect the cluster size at which the transition from external to internal state of the excess electron occurs, allowing it to happen at smaller cluster sizes when they are included. Therefore, this work reinforces the importance of including nuclear quantum effects to obtain a precise picture when modelling water cluster in the presence of an excess electron, apart from employing a suitable water-electron interaction.

\section{Acknowledgments}

We thank J. M. Soler, M. V. Fern\'{a}ndez-Serra, R. Ram\'{i}rez and C. P. Herrero for helpful discussions. This work has been supported by the Spanish State Agency for scientific research through project PGC2018-096955-B-C44.

\appendix
\section{Computational details}
\label{sec:appendix}

\subsection*{Simulation procedure}
\label{sub:simulation}

One of the central aims of this work is to account for the quantum effects of the atomic dynamics in the structural and energetic properties of water clusters, and especially those of water cluster anions, where we anticipate that such effects could be most appreciable, particularly on the excess electron properties. To do so, we employ the Feynman Path Integral~(PI) approach. As is well known, the PI formalism establishes a quantum-classical isomorphism, through which the partition function of a quantum system can be approximately mapped onto that of a classical system consisting of $P$ replicas (also referred to as {\em beads}) of the original system, but subject to a modified potential, as follows:
\begin{widetext}
\begin{equation}
  Q_P =  \left[ \prod_{i=1}^N \left( \frac{m_i P}{2\pi\beta\hbar^2} \right) ^{3P/2} \int
  d\bm{r}_i^1 \cdots d\bm{r}_i^P \right]
   \times 
  \exp{ \left\{ -\beta \sum_{\alpha=1}^P \left[ \sum_{i=1}^N \frac{m_i P}{2\beta^2\hbar^2}( \bm{r}_i^{\alpha}
  - \bm{r}_i^{\alpha+1} )^2
  + \frac{V(\bm{r}^{\alpha})}
  {P} \right] \right\}
  }_{\bm{r}_i^{P+1}=\bm{r}_i^1}.
 \label{eq:partition}
\end{equation}
\end{widetext}
Here, $Q_P$ is the partition function of the {\em P\/}-bead classical isomorph approximating that of its quantum counterpart; $\beta = (k_B T)^{-1}$, and $V(\bm{r}^\alpha)$ is the potential energy. Latin sub-indices label atoms, while Greek super-indices label beads; if only a super-index is given, the variable is meant to represent the entire configuration for that bead, i.e. $\bm{r}^\alpha \equiv (\bm{r}_1^\alpha, \bm{r}_2^\alpha \ldots \bm{r}_n^\alpha)$. The isomorphism is exact in the limit $P\rightarrow \infty$. For a finite number of replicas, thermodynamic estimators for the internal energy $E$, its kinetic and potential contributions, and a series of other properties can be easily derived~\cite{Gillan1990,Tuckerman2010,Herman1982} from the partition function Eq.~(\ref{eq:partition}). Obtaining average equilibrium properties of the quantum system then reduces to sampling the classical isomorph in the canonical ensemble, either with Monte Carlo or with Molecular Dynamics techniques.

In the case of the neutral clusters $V(\bm{r})$ in Eq.~(\ref{eq:partition}) reduces to $V_{\text{SPC/F}}(\bm{r})$, i.e. the SPC/F potential energy expression. However, for the the cluster anions the interaction of water molecules with the excess electron must be incorporated into the model. Next we describe how we have done this in our simulations.

\subsection*{Schr\"{o}dinger equation for the excess electron}
\label{sub:schroedinger}

In principle it would be possible to address the dynamics of water molecules and excess electron on an equal footing within the PI formalism. However, the light mass of the electron would require a large number of beads in the PI ring polymer, much larger than would be required in the absence of the electron. While it is possible to effectively use different numbers of beads for different degrees of freedom~\cite{Thirumalai1986,Wallqvist1987}, we have found it more convenient to invoke the Born-Oppenheimer approximation and treat separately the dynamics of molecules and excess electron. Thus we use the PI method to account for the quantum effects in the dynamics of the water molecules subject to their mutual interaction and that of the electron density charge, while the latter is obtained for each cluster configuration by direct numerical solution of the electron's Schr\"{o}dinger equation.

To obtain the electron charge distribution we proceed as follows: first, the interaction potential of the electron with the water molecules, Eq.~(\ref{eq:w-e-potential}), is discretized on a regular cubic grid covering a volume large enough to contain the water cluster, with its center fixed at the cluster center of mass. In this work we have used a grid of size $L=80$\,\AA\ with a grid of $64\times 64 \times 64$ grid points, which provides sufficient resolution, as we discuss below in the convergence tests subsection. The lowest~4 electron eigen-pairs are then obtained iteratively using the imaginary time propagation method~\cite{Chin2009}. In this method, a set of {\em n\/} trial wave functions $\psi_i$ is driven towards the {\em n\/} lowest lying eigenstates of Hamiltonian $H$ by acting on them with the imaginary time evolution operator:
\begin{equation}
    \psi_i(\epsilon) = e^{-\epsilon H} \psi_i(0).
    \label{eq:evolution}
\end{equation}
Here $\epsilon$ is the imaginary time-step.
Notice that the complex time evolution operator is not unitary, and hence does not preserve the orthonormality of the trial set; it is therefore necessary to orthonormalise the set following each time-step. After a sufficiently large number of evolution steps (followed by orthonormalisation), this process converges to the lowest {\em n\/} eigenstates of $H$. Once these eigenstates have been obtained, the excess electron charge density is obtained 

The action of operator $e^{-\epsilon H}$ on the trial set cannot be calculated exactly, but it can be approximated up to any desired order in $\epsilon$~\cite{Chin2009} using a Trotter-like factorisation:
\begin{equation}
    e^{-\epsilon H} \approx \sum_k c_k \prod_j e^{-a_{jk} T} e^{-b_{jk} V},
    \label{eq:trotter}
\end{equation}
where $c_k, a_{jk}$, and $b_{jk}$ are numerical coefficients, and $H = T + V$, where $T$ and $V$ are the kinetic and potential energy operators, respectively. Operators $e^{\alpha T}$ are diagonal in momentum space, while operators $e^{\beta V}$ are diagonal in real space (provided $V$ is local, as is the case in our model), so the action of Eq.~(\ref{eq:trotter}) on the trial set involves a succession of Fast Fourier transformations~\cite{FFTW97} between momentum and real space (see Ref.~\onlinecite{Chin2009} for details). In practice we find that a $4^{th}$-order approximation to $e^{-\epsilon H}$ provides a good compromise between computational effort and speed of convergence. 

In the case of the cluster anions $V(\bm{r})$ in Eq.~(\ref{eq:partition}) takes the form:
\begin{equation}
    V(\bm{r}^\alpha) = E_e(\bm{r}^\alpha) + V_{\text{SPC/F}}(\bm{r}^\alpha),
\label{eq:ionpotential}
\end{equation}
where $E_e(\bm{r}^\alpha)$ is the electron ground state energy for bead $\alpha$, which  depends parametrically on $\bm{r}^\alpha$, and $V_{\text{SPC/F}}(\bm{r}^\alpha)$ is the SPC/F model potential for water, as in the neutral clusters. 

\subsection*{Centroid approximation}
\label{sub:centroid}

Strictly speaking, the ground-state energy, $E_e(\bm{r}^{\alpha})$, must be evaluated separately for each replica $\alpha$, and then accumulated with weight $1/P$. However, as argued below, it is possible to take a further approximation in our scheme, which we refer to as the {\em centroid approximation\/}. 
Each quantum atomic nucleus will be delocalized with certain mean position and spatial extension, characterized by its centroid (instantaneous center of the chain) and its gyration radius:
\begin{align}\label{eq:centroid-def}
  \bar{\bm{r}}_i & = \frac{1}{P}\sum_{\alpha=1}^P \bm{r}_i^\alpha \, , \\
  r_{g,i}^2 & = \frac{1}{P} \sum_{\alpha=1}^P ( \bm{r}_i^\alpha
  - \bar{\bm{r}}_i )^2 \, . \label{eq-Rg}
\end{align}
The centroid approximation amounts to solving the electron Schr\"{o}dinger equation for the ring polymer centroid instead of doing it for each individual bead.
\begin{figure}[t]
\centering 
\includegraphics[scale=1.4]{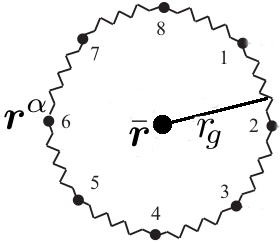}
\hspace{0.5cm}
\includegraphics[scale=0.3]{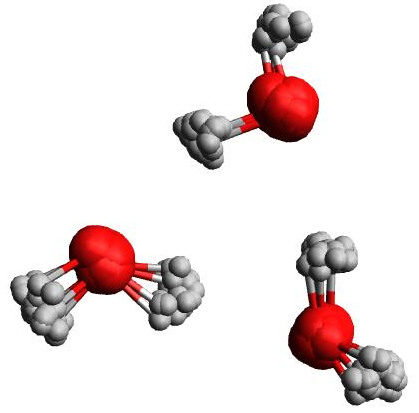}
\caption{(Left) Schematic representation of the beads, $\bm{r}^{\alpha}$, centroid position, $\bar{\bm{r}}$, and gyration radius, $r_g$, of a classical ring polymer corresponding to a quantum particle with 8 replicas. (Right) Typical geometry of a 3 water molecules cluster, simulated with 40 beads.}
\label{fig:ring-polymer}
\end{figure}

A schematic representation of a quantum particle with 8 beads around its centroid position can be seen in left panel of Fig.~(\ref{fig:ring-polymer}). 

Like the energy, the forces on the molecular atoms in the cluster anion simulations have two contributions: one resulting from the water-water interactions, modelled via the SPC/F model, and the second from their interaction with the excess electron charge density, which can be easily evaluated from the Hellman-Feynman theorem. Let us consider the electron energy at bead $\alpha$ and write it in terms of that evaluated for the centroid; we would have:
\begin{equation}\label{eq:centroid-expansion}
  E_e( \bm{r}^{\alpha} ) = 
  E_e( \bar{ \bm{r} } )
  + \nabla E_e( \bar{ \bm{r} } ) \cdot ( \bm{r}^{\alpha} - \bar{ \bm{r} } )
  + \mathcal{O}( |\bm{r}^{\alpha} - 
  \bar{ \bm{r} }|^2 ) \,.
\end{equation}
The centroid approximation implies taking $E_e( \bm{r}^{\alpha} ) \approx E_e( \bar{ \bm{r} } )$, which will be justified when the second and subsequent terms on the rhs of Eq.~(\ref{eq:centroid-expansion}) are negligible compared to $E_e( \bar{ \bm{r} } )$. This will happen if the Hellman-Feynman forces are small and/or the gyration radius is small. Typically, the distances between bead coordinates of the same quantum ion are much smaller than the internuclear distances, as can be appreciated in the right panel of Fig.~(\ref{fig:ring-polymer}), and so we can expect the approximation to hold valid.
 Within this approximation, the electronic contribution to force acting on the $i$th atom in replica $\alpha$ due to the electron cloud is given by
\begin{equation}
    \bm{F}_{e,i}^\alpha = \frac{1}{P} \bar{\bm{F}}_{e,i} \, ,
\end{equation}
as can be seen using the chain rule, 
where $\bar{\bm{F}}_{e,i}$ is the electronic force acting on the centroid of atom $i$.

From a practical point of view, the centroid approximation reduces considerably the computational cost needed to obtain the electronic energy and forces, as this is done only for the centroid configuration and not for every bead on the ring polymer. Nevertheless, the approximation needs to be justified by comparing its predictions against those of the rigorous procedure, which will be done later.

\subsection*{Molecular Dynamics sampling}
\label{sub:MD}
In order to sample the partition function Eq.~(\ref{eq:partition}) and obtain thermal averages derived from it we employ molecular dynamics simulations at constant temperature. To ensure sampling of the canonical ensemble we combine the ring polymer dynamics with a Bussi-Parrinello~\cite{Bussi2007} thermostat attached to each degree of freedom. We used a time step of 1\,fs and a friction parameter $\gamma = 0.001$\,au, which provided effective sampling.

For classical simulations of the neutral clusters, initial geometries were obtained from large molecular dynamics simulations of bulk water at 300\,K and density 1\,$\text{g/cm}^3$, including all water molecules within spheres of varying radii~\cite{Rudberg2018}. For each calculation, an equilibration run of $1\times10^5$ steps (100\,ps) was executed before the production simulation of $5\times10^5$ steps. Quantum PIMD simulations were carried out with 128 replicas for all temperatures between 50 and 400\,K, using classical equilibrated configurations as initial conditions. As will be justified later, the election of $P=128$ provides converged properties and it is computationally affordable for neutral clusters.  

For water cluster anions, the simulations were initiated from interior states, obtained from neutral configurations with a relaxation simulation in the presence of a smooth confining potential to keep the excess electron in the vicinity of the cluster center, $V_{\text{conf}}=\frac{1}{2}k(x^8 + y^8 + z^8)$, with $k=10^{-8}$\,au. The relaxation was performed at each temperature during $10^4$ steps, and after that the confining term was switched off. Typical calculations consisted of an equilibration run of $5\times10^4$ steps (50\,ps), previous to the production run with length between $10^5$ and $2\times10^5$ steps, which correspond to 100 and 200\,ps of time. For the quantum simulations of the negatively charged water clusters, we used 128, 64 and 48 replicas for temperatures of 100, 200 and 300\,K, respectively, trying to keep constant the product $PT$ as close as possible. In addition, we employed the centroid approximation, which provides accurate results for both the water cluster and the excess electron properties and its computational needs are up to two orders of magnitude lower than the exact calculation at 100\,K. The election of those number of replicas and the use of the centroid approximation is justified below in the next subsection.

In all cases, the associated uncertainties of equilibrium averages were computed with block averaging, as the plateau of the standard error of the mean among the blocks, when increasing the block size~\cite{Allen2017,Grossfield2018}. 

\subsection*{Convergence tests}
\label{sub:convergence}

\begin{figure}[t]
    \centering
    \includegraphics[scale=0.35]{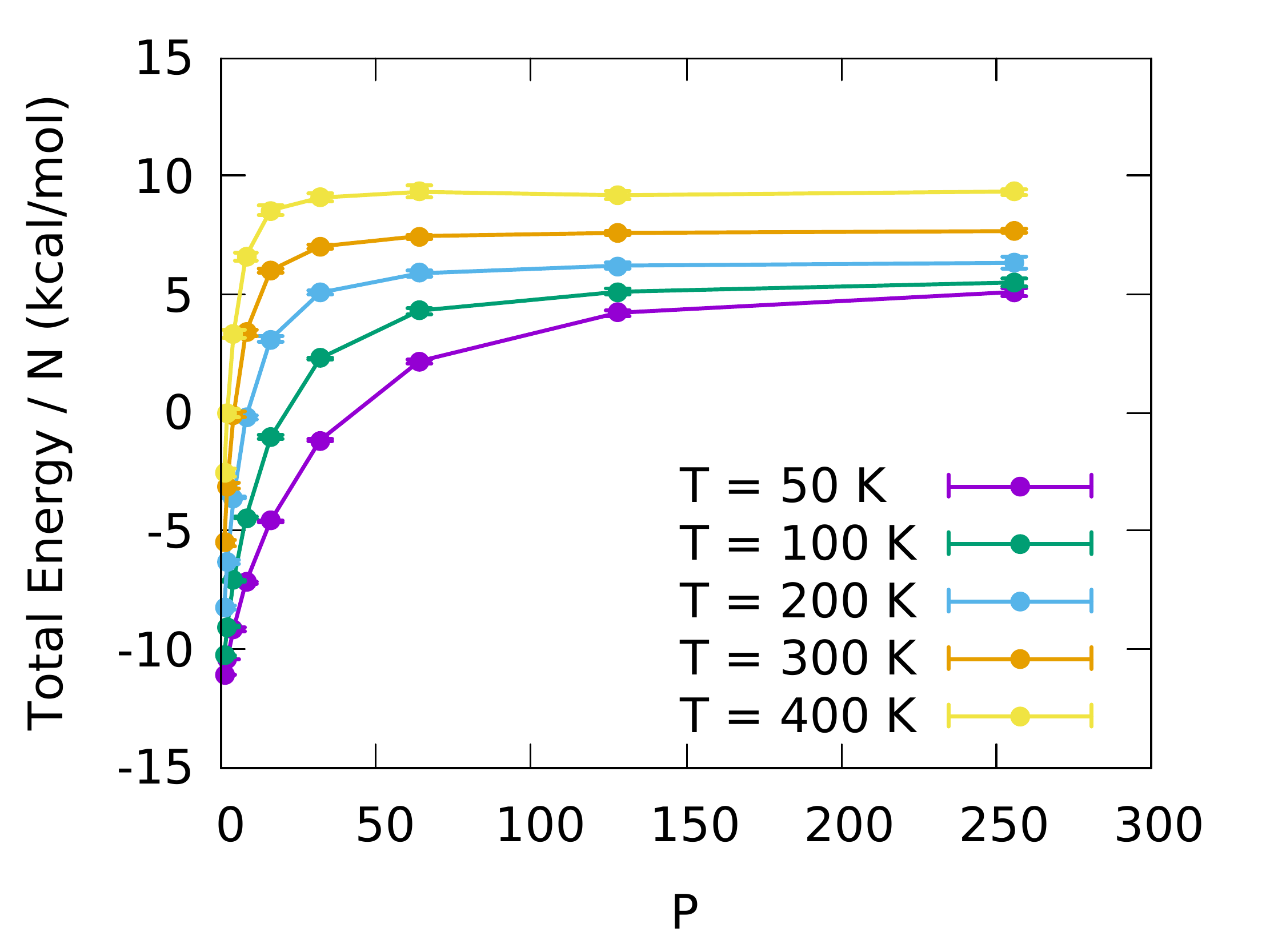}
\caption{Total energy per molecule as a function of the number of replicas, for several temperatures.}
\label{fig:convergence}
\end{figure}

\begin{table}[tbh]
    \begin{center}
    {\normalsize{}Surface state, $N$=100, $T=300\,$K}{\large\par}
    \par\end{center}
    \begin{tabular}{|c|c|c|c|c|c|c|}
    \cline{2-7} 
    \multicolumn{1}{c|}{} & \multicolumn{2}{c|}{$N_{p}=32$} & \multicolumn{2}{c|}{$N_{p}=64$} & \multicolumn{2}{c|}{$N_{p}=128$}\tabularnewline
    \cline{2-7} 
    \multicolumn{1}{c|}{} & $\varepsilon_{0}$ & $\varepsilon_{1}$ & $\varepsilon_{0}$ & $\varepsilon_{1}$ & $\varepsilon_{0}$ & $\varepsilon_{1}$\tabularnewline
    \hline 
    $L=40$ & -1.2596 & -0.4268 & -1.2393 & -0.4009 & -1.2285 & -0.3872\tabularnewline
    \hline 
    $L=80$ & -1.3454 & -0.5995 & -1.3560 & -0.5944 & -1.3560 & -0.5944\tabularnewline
    \hline 
    $L=160$ & -1.5857 & -0.8469 & -1.3511 & -0.6025 & -1.3561 & -0.5961\tabularnewline
    \hline 
    \end{tabular}
    \caption{Ground and first excited state energies of the excess electron as a function of the grid size $L$ and the number of grid points $N_p$, in each Cartesian direction. All the calculations correspond to the $n=100$ neutral cluster. Energies are in eV and lengths in \AA.}
    \label{table:convergence}
\end{table}

Here we analyze the convergence of the total energy with the number of replicas for a cluster with $N$=100 water molecules at different temperatures. As anticipated, the number of replicas required to converge increases with decreasing temperature, as can be observed in the top panel of Fig.~(\ref{fig:convergence}). While $P=32$ is sufficient at $T=400$\,K, a similar level of convergence at $T=50$\,K requires 256 replicas. The differences between using 128 and 256 replicas are approximately 5\,\%\ or less for temperatures equal to or higher than 100~K. Consequently, we performed all subsequent simulations at a fixed $PT=12800$\,K value, which implies using 128 replicas at $T=100$\,K, 64 at $T=200$\,K, and 48 at $T=300$\,K.

\begin{figure}[t]
\centering 
\includegraphics[scale=0.4]{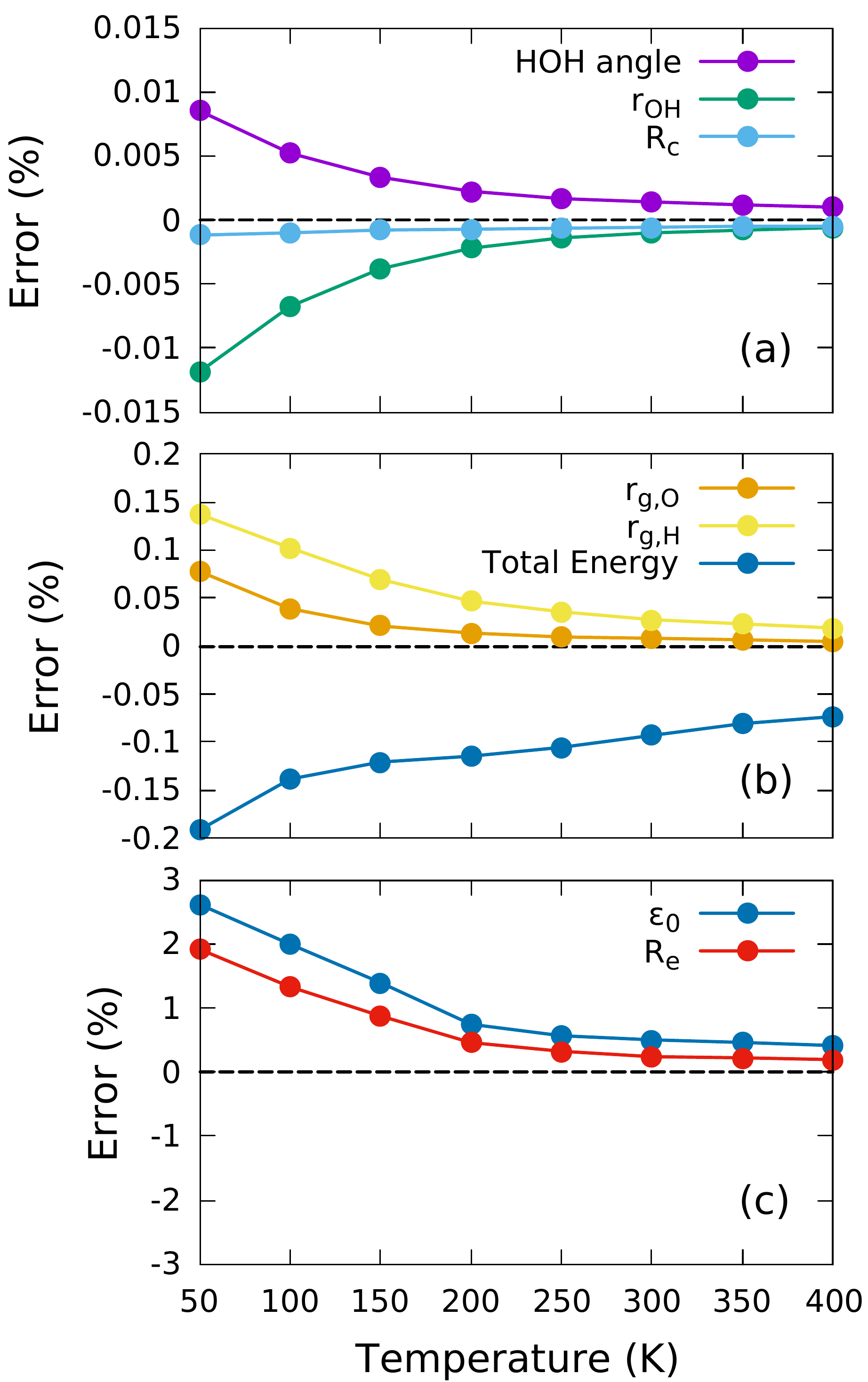}
\caption{Relative error made in the averages when using the centroid approximation to study the $(\text{H}_2\text{O})_{100}^{-}$ cluster, with respect to the exact quantum simulation. Panel (a) shows the error in the intramolecular HOH angle and the OH distance, and the estimated cluster radius. Panel (b) displays the error in the total energy of the system and the gyration radius of oxygen and hydrogen atoms. Lastly, panel (c) presents the differences in the excess electron ground-state energy, $\varepsilon_0$, and its spatial width, $R_e$.}
\label{fig:centroid-error}
\end{figure}

For cluster anions we also require a suitable grid with appropriate size and grid spacing on which to discretize the electron wave functions and density. For this purpose we analysed the convergence of the ground and first excited state energies for an electron in a surface state in the $n=100$ cluster; interior states are more localized and show better convergence with the grid parameters. The results, obtained with the complex time evolution using an energy tolerance of $10^{-4}\,\text{Hartree} \approx 2.7\times10^{-3}$\,eV, are shown in  Table~(\ref{table:convergence}). As can be seen there, a cubic grid extending over 80\,\AA\ and with 64~points along each dimension is capable of providing converged eigenenergies with an error smaller than $10^{-2}$~eV. A grid twice as large in volume but with equal grid point density provides essentially indistinguishable results. Therefore, for all subsequent simulations of cluster anions we employed $L=80$\,\AA\ and $N_p=64$ for all sizes and temperatures.

As discussed above, the centroid approximation allows us to reduce the computational cost of the simulations, but its validity must be checked by comparing its predictions against those of the rigorous procedure. To do so, we have performed 20~ps-long simulations ($2\times 10^4$ time steps) for the $n=100$ cluster anion at different temperatures with 128 replicas, starting from previously equilibrated configurations. In Fig.~(\ref{fig:centroid-error}) we plot the relative differences in electron energy and spread, defined as
\begin{equation}
  R_e = \left[ \int | \bm{r} - \bar{\bm{r}} |^2 \rho( \bm{r} ) d\bm{r} \right]^{1/2} \,,
  \label{eq:egyration}
\end{equation}
where $\rho$ is the electron density, between the simulations with and without centroid approximation. The relative difference decreases from 2.5 to 0.5\,\%\ between temperatures from 50 to 400~K, due to the fact that the kinetic energy spring constant increases with temperature ($\propto T^2$), resulting in more localized beads around the centroid. For structural and energetic properties of the entire cluster anion the errors are even smaller ($< 0.5$\,\%). 

As can be seen in Table~(\ref{table:convergence}), the energy gap between ground and first excited state of the excess electron in cluster anions remains significantly larger than $k_\text{B} T$ even at the largest temperature considered for these systems (300 K); thus it is safe to assume that the electron remains in its ground state throughout. 

\bibliographystyle{apsrev4-1}
\bibliography{./References.bib}
\nocite{*}

\end{document}